\documentclass[a4paper,twocolumn,11pt,accepted=9999-99-99]{quantumarticle}
\pdfoutput=1
\usepackage[utf8]{inputenc}
\usepackage[english]{babel}
\usepackage[T1]{fontenc}
\usepackage{amsmath}
\usepackage{amssymb}
\usepackage{hyperref}

\usepackage{tikz}
\usepackage{braket}
\usepackage{lipsum}
\usepackage{quantikz}
\usepackage{subfigure}

\usepackage[numbers,sort&compress]{natbib}

\begin{document}

\title{Fast Expectation Value Calculation Speedup of Quantum Approximate Optimization Algorithm: HoLCUs QAOA}

\author{Alejandro Mata Ali}
\affiliation{Instituto Tecnológico de Castilla y León, Burgos, Spain}
\email{alejandro.mata@itcl.es}
\orcid{0009-0006-7289-8827}

\maketitle

\begin{abstract}
In this paper, we present a new method for calculating expectation values of operators that can be expressed as a linear combination of unitary (LCU) operators. This method allows to perform this calculation in a single quantum circuit measuring a single qubit, which speeds up the computation process. This method is general for any quantum algorithm and is of particular interest in the acceleration of variational quantum algorithms, both in real devices and in simulations.

We analyze its application to the parameter optimization process of the Quantum Approximate Optimization Algorithm (QAOA) and the case of having degenerate values in the matrix of the Ising problem. Finally, we apply it to several Quadratic Unconstrained Binary Optimization (QUBO) problems to analyze the speedup of the method in circuit simulators.
\end{abstract}

\section{Introduction}
Combinatorial optimization~\cite{Combinatorial} is one of the fields with the greatest applicability to real-world problems. However, many combinatorial optimization problems of industrial interest are NP-Hard problems. This implies that their exact classical resolution scales exponentially with the instance size. This limits the possibility of using classical algorithms to solve them in a reasonable amount of time. Examples are the traveling salesman problem~\cite{TSP} or the knapsack problem~\cite{Knapsack_original}. This has motivated the creation of several approximate algorithms~\cite{Aproximated,Genetic,Heuristics} with the aim of offering enough good solutions, even if they are not optimal. An example are the route problems, in which a company may not be interested in obtaining the best route if it can obtain a slightly worse one in a much shorter time.  In this context, interest arises in combinatorial optimization using quantum algorithms.

The field of quantum computing has gained interest in recent years because of its application to various problems that are difficult for classical computing. An example are quantum algorithms with a proven speedup with respect to their classical counterparts, such as Shor's algorithm~\cite{Shor} or the quantum linear solver HHL~\cite{HHL}. There are also several quantum algorithms that have been used to solve combinatorial optimization problems~\cite{Quantum_Optim,Quantum_Block_Optim,Generative_Quantum_Combinat,Bench_Quantum_Optim,HOBO_Quantum_2,Variational_Quantum_Optim,VQC_HOBO,VRP_Quantum}. However, given the current state of quantum hardware, in the Noisy intermediate-scale quantum (NISQ) era~\cite{NISQ}, most of these algorithms are not suitable on quantum devices. In this context, new types of algorithms emerge, NISQ-friendly~\cite{NISQ_Algor}, that aim to run on current devices and offer an advantage. These algorithms are able to work in spite of noise, either by low sensitivity to it or by absorbing it in their variational parameters.

One of the most popular quantum algorithms is the Quantum Approximate Optimization Algorithm (QAOA)~\cite{QAOA}. As its name suggests, it is an approximate algorithm dedicated to combinatorial optimization. This is a NISQ-friendly variational algorithm that seeks to map the combinatorial problem to a hamiltonian, and find the solution to it by finding the ground state of this hamiltonian. To do this, it has a specially designed ansatz, with parameters that we must adjust variationally until the quantum circuit returns the correct result with maximum probability.

However, the tuning process requires minimizing the energy of the system, which implies determining it at each tuning step. Since the problem hamiltonians are usually not unitary operators, but hermitian, this computation cannot be done directly by means of a Hadamard test. Although we can measure the expectation values of some hermitic operators, usually for general problems we need to measure several separately and sum them weighted. This may inappropriately increase the circuit parameters training time. This problem does not appear only in the case of QAOA, but is common for a large number of variational quantum algorithms~\cite{Variational_Quantum_Optim}. Previous work has presented methods to obtain the expectation value of a linear combination of unitaries (LCU), but requiring the measurement of several qubits~\cite{Old_H_LCU}.

In this paper, we present an algorithm to calculate the expectation value $\braket{\psi|A|\psi}$, given an input quantum state $\ket{\psi}$ and an operator $A$ that can be expressed as an LCU, with a single circuit and measuring a single qubit, adding a logarithmic amount of extra qubits. For this purpose, we combine the Hadamard test with the protocol for applying LCU to a quantum state. Although the LCU protocol is probabilistic and may fail, our protocol is deterministic and does not fail. We call this method \textit{Hadamard+LCU Test}, \textit{H-LCU} or \textit{HoLCUs}, in reference to the plant. Although it is a general method applicable to any quantum circuit, it is especially interesting for training variational circuits. To see its potential, we are going to particularize it in the case of the QAOA, since it is the algorithm in which it is simplest to understand. In addition to offering a considerable speedup compared to the conventional Hadamard test, the speedup is superior in cases of high degeneracy of the LCU coefficients.  We will implement the QAOA with HoLCUs incorporated in Qiskit and compare it with the QAOA without HoLCUs obtaining the expectation values separately. The code is available in the repository \href{https://github.com/DOKOS-TAYOS/HoLCUs_QAOA}{https://github.com/DOKOS-TAYOS/HoLCUs\_QAOA}.

\section{Description of the QAOA}

First, we will briefly describe the QAOA algorithm. This algorithm is designed to obtain approximate solutions to combinatorial optimization problems. In this way, we can map the cost function of the problem into a hamiltonian, so the highest-cost states have higher energy. Its objective is, given a problem hamiltonian $H_P$, to return its quantum state of minimum energy, its ground state. For this purpose, it makes use of a variational quantum circuit composed of two main parts. The first part implements the time evolution of the state with the problem hamiltonian $H_P$. That is, this part of the circuit implements the operator
\begin{equation}
    U_P = e^{i\gamma H_P},
\end{equation}
where $\gamma$ is the variational parameter we need to adjust. This part associates to each eigenstate of the problem hamiltonian a phase proportional to its energy.

The second part of the circuit consists of a circuit that implements the time evolution with a mixing hamiltonian $H_M=\sum_k \sigma^X_k$. Therefore, the operator applied to the state is
\begin{equation}
    U_M = e^{i\beta H_M},
\end{equation}
where $\beta$ is the parameter to be adjusted. This part is in charge of making the phases of the eigenstates interfere, bringing amplitude to the lower-energy eigenstates. 

The concatenation of these two parts results in a QAOA layer. Therefore, in the complete QAOA circuit with $p$ layers, we initialize the state in uniform superposition and apply the $U_P$ and $U_M$ operators $p$ times. This way, the resulting $n$ qubits state is
\begin{equation}
    \ket{\psi} = QAOA\ket{0}^{\otimes n} = \prod_{j=0}^{p-1}U_M(\beta_{j})U_P(\gamma_{j})\ket{+},
\end{equation}
being $\ket{+}=\frac{1}{\sqrt{2^{n}-1}}\sum_{k=0}^{2^{n}-1}\ket{k}$ the uniform superposition.

Therefore, to obtain the desired output, we must adjust the $\vec{\gamma}$ and $\vec{\beta}$ angles so that the expectation value $\braket{\psi|H_P|\psi}$ is minimized.
The implementation of the $U_M$ operator is done by making use of $RX(\beta_j)$ gates in each of the qubits. The implementation of the $U_P$ operator depends on the problem, but usually focuses on problems that can be expressed as a linear combination of hamiltonians $H_k$
\begin{equation}
    H_P=\sum_k \alpha_k H_k.
\end{equation}

A case of great interest is the case of Quadratic Unconstrained Binary Optimization (QUBO)~\cite{QUBO} problems, which can be mapped to Ising problems~\cite{Ising}, so that the hamiltonian of the problem is of the type
\begin{equation}
    H_P = \sum_k \alpha_k \sigma^Z_k + \sum_{k,l} \alpha_{k,l} \sigma^Z_{k}\otimes \sigma^Z_{l}.
\end{equation}

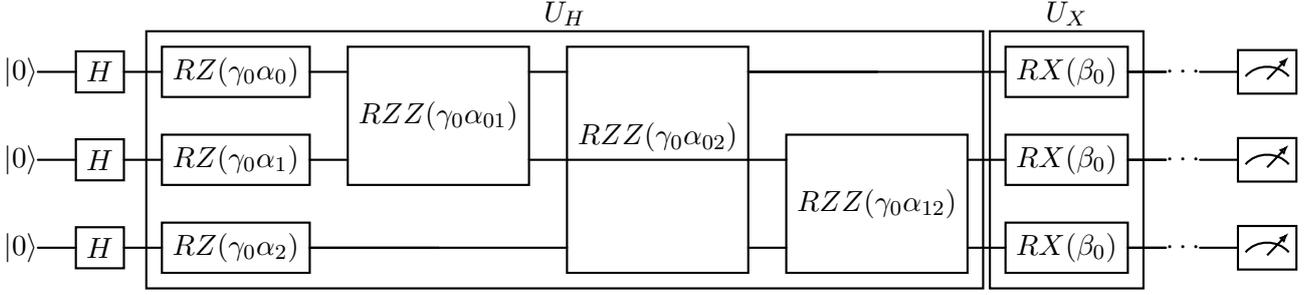
\begin{figure*}
    \centering
    \begin{quantikz}[transparent]
    \ket{0} & \gate{H} & \gate{RZ(\gamma_0\alpha_0)}\gategroup[3,steps=4,style={inner sep=2pt}]{$U_H$}& \gate[2]{RZZ(\gamma_0\alpha_{01})} & \gate[3,label style={yshift=0.3cm}]{RZZ(\gamma_0\alpha_{02})} & \qw& \gate{RX(\beta_0)}\gategroup[3,steps=1,style={inner sep=2pt}]{$U_X$} &\qw \cdots & \meter{}\\
    \ket{0} & \gate{H} & \gate{RZ(\gamma_0\alpha_1)} &  & \linethrough & \gate[2]{RZZ(\gamma_0\alpha_{12})} & \gate{RX(\beta_0)} & \qw\cdots & \meter{}\\
    \ket{0} & \gate{H} & \gate{RZ(\gamma_0\alpha_2)}& \qw &  & & \gate{RX(\beta_0)} & \qw\cdots & \meter{}
    \end{quantikz}
    \caption{QAOA circuit with one layer and three qubits.}
    \label{fig: QAOA circuit}
\end{figure*}

In this case we can express the operator $U_P$ as a product of gates $RZ(\gamma_j\alpha_{k})$ and $RZZ(\gamma_j\alpha_{k,l})$. An example is shown in Fig.~\ref{fig: QAOA circuit}.

%%%%%%%%%%%%%%%%%%%%%%%%%%%%%%%%%%%%%%%%%%%%
%%%%%%%%%%%%%%%%%%%%%%%%%%%%%%%%%%%%%%%%%%%%
%%%%%%%%%%%%%%%%%%%%%%%%%%%%%%%%%%%%%%%%%%%%
%%%%%%%%%%%%%%        H+LCU     %%%%%%%%%%%%
%%%%%%%%%%%%%%%%%%%%%%%%%%%%%%%%%%%%%%%%%%%%
%%%%%%%%%%%%%%%%%%%%%%%%%%%%%%%%%%%%%%%%%%%%
%%%%%%%%%%%%%%%%%%%%%%%%%%%%%%%%%%%%%%%%%%%%
%%%%%%%%%%%%%%%%%%%%%%%%%%%%%%%%%%%%%%%%%%%%
\section{Expectation value estimation algorithm HoLCUs}
To perform the QAOA train process, we need to obtain the expectation value of $H_P$ from the output state of the QAOA circuit at each train step. This usually consists of determining the expectation value of a linear combination of unitaries. For this we have 3 alternatives. The first is to directly measure the operator $H_P$. However, in general cases, this could not be possible. The second is to know the final statevector and compute $H_P$ by brute force, but this would be as expensive as solving the problem by brute force. The third one is to apply the Hadamard test to each Pauli-string and sum the results obtained by each one.

The disadvantages of this method are:
\begin{itemize}
    \item The increase of the number of terms of the hamiltonian to be evaluated with the number of variables, which implies running the circuit a number of times that may be undesirable. In QUBO problems, this scaling is quadratic.
    \item The different scale of the different terms of the hamiltonian, for which we have to increase the number of shots in some terms with respect to others so that the statistical deviation of the larger terms does not hide the smaller terms.
\end{itemize}

To avoid these problems, we decided to unify the Hadamard Test to obtain the expectation values of a unitary operator with a method that allows the nondeterministic application of a non-unitary operator given by the linear combination of unitary operators (LCU). We call this new technique \textit{Hadamard+LCU Test}, \textit{H-LCU} or \textit{HoLCUs}.

There are already methods in the literature that study how to perform such an operation~\cite{Old_H_LCU}, but we propose a new method that allows the measurement of a single qubit and is a deterministic operation. For this purpose we will briefly introduce both the Hadamard Test and the LCU method.

\subsection{Hadamard Test}
The Hadamard test is a widely known technique for the calculation of expectation values of unitary operators, consisting in the use of interference to measure only one ancilla qubit and obtain the expectation value.

\begin{figure}
\centering
\begin{quantikz}[column sep=5pt, row sep={20pt,between origins}]
  \lstick{$\ket{0}_a$} \slice{1}& \gate{H}\slice{2} & \gate[style={blue!20}]{S^\dagger} & \ctrl{1} \slice{3}& \gate{H} \slice{4}& \meter{} \\
  \lstick{$\ket{\psi}$} & \qw & \qw & \gate{U} & \qw & \qw  
\end{quantikz}
\caption{Quantum circuit for the Hadamard Test of state $|\psi\rangle$ and a unitary operator $U$. The $S^\dagger$ gate will be on if we want the imaginary part of the expectation value and off if we want the real part.}
\label{cirq: Hadamard}
\end{figure}
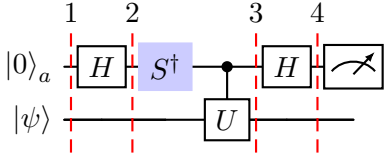

Given a state $|\psi\rangle$ and a unitary operator $U$, we want to compute the expectation value of $\langle\psi|U|\psi\rangle$. For this we use the circuit of Fig. \ref{cirq: Hadamard}, where we have one qubit ancilla $|0\rangle_a$.

The evolution process of the system along the quantum circuit, without the $S^\dagger$ gate, is as follows.
\begin{enumerate}
    \item $\ket{0}_a \ket{\psi}$
    \item $\frac{\ket{0}_a+\ket{1}_a}{\sqrt{2}}\ket{\psi}$
    \item $\frac{\ket{0}_a}{\sqrt{2}}\ket{\psi}+\frac{\ket{1}_a}{\sqrt{2}}U\ket{\psi}$
    \item $\frac{\ket{0}_a}{2}\left(\ket{\psi}+U\ket{\psi}\right) + \frac{\ket{1}_a}{2}\left(\ket{\psi}-U\ket{\psi}\right)$
\end{enumerate}

So the probability $P(0)$ of measuring $\ket{0}$, or $P(1)$ of measuring $\ket{1}$, in the qubit ancilla is
\begin{align}
    P(0)=\frac{1}{2}\left(1+Re[\bra{\psi}U\ket{\psi}]\right),\\
    P(1)=\frac{1}{2}\left(1-Re[\bra{\psi}U\ket{\psi}]\right),
\end{align}
so that
\begin{equation}
    Re[\bra{\psi}U\ket{\psi}] = 2P(0)-1.
\end{equation}
For the imaginary part, we only have to activate the $S^\dagger$ gate, so that we have
\begin{equation}
    Im[\bra{\psi}U\ket{\psi}] = 2P(0)-1.
\end{equation}

\subsection{Linear Combination of Unitaries}
Given a state $\ket{\psi}$ and an operator $A$,  we want to apply the operation $A\ket{\psi}$. The operator $A$ can be expressed as a linear combination of $M$ unitary operators
\begin{equation}
    A=\sum_{k=0}^{M-1} \alpha_k \xi_k U_k,
\end{equation}
being $\alpha_k\in \mathbb{R}>0$, $\xi_k=e^{i\theta_k}$ and $U_k$ is a unitary operator, typically a Pauli-string. It must respect the normalization restriction
\begin{equation}
    \sum_{k=0}^{N-1} \alpha_k = 1.
\end{equation}

To perform this operation, we use a set of $m = \lceil \log_2(M)\rceil$ ancilla qubits. We define the operators
\begin{align}
    B(\psi)\ket{0}& = \ket{\psi},\\
    V\ket{0}& = \sum_{k=0}^{M-1}\sqrt{\alpha_k}\xi_k\ket{k},\\
    \hat{V}\ket{0}& = \sum_{k=0}^{M-1}\sqrt{\alpha_k}\ket{k}.
\end{align}

\begin{figure}
\centering
\begin{quantikz}[column sep=5pt, row sep={20pt,between origins}]
  \lstick{$\ket{0}_a$} \slice{1}& \gate[2]{V}\slice{2} & \ctrl[open]{1} & \ctrl[open]{1} & \ctrl{1}       & \ctrl{1}\slice{3} & \gate[2]{\hat{V}^\dagger} \slice{4}& \qw \\
  \lstick{$\ket{0}_a$} &             & \ctrl[open]{1} & \ctrl{1}       & \ctrl[open]{1} & \ctrl{1} &                             & \qw \\
  \lstick{$\ket{0}$}   & \gate[2]{B(\psi)} & \gate[2]{U_0} & \gate[2]{U_1} & \gate[2]{U_2} & \gate[2]{U_3} & \qw & \qw \\
  \lstick{$\ket{0}$} &  &  &  &  &  & \qw & \qw
\end{quantikz}
\caption{Quantum circuit for the LCU of a 2-qubits state $|\psi\rangle$ and an operator $A=\sum_{k=0}^{3} \alpha_k \xi_k U_k$, with 2 ancilla qubits.}
\label{cirq: LCU}
\end{figure}
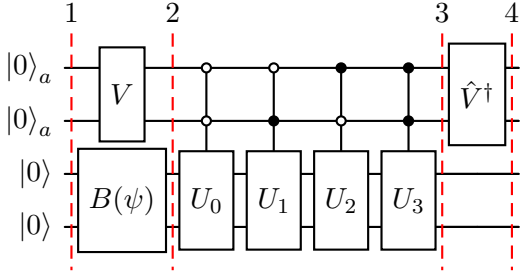
Following the circuit shown in Fig. \ref{cirq: LCU}, the state evolves as follows
 \begin{enumerate}
     \item$ \ket{0}_a^{\otimes m} \ket{0}$
     \item$(V\otimes B(\psi))\ket{0}_a^{\otimes m} \ket{0} = \sum_{k=0}^{M-1}\sqrt{\alpha_k}\xi_k\ket{k}_a\ket{\psi}$
     \item$\sum_{k=0}^{N-1}\sqrt{\alpha_k}\xi_k(\mathbb{I}\otimes U_k)\ket{k}_a\ket{\psi}$
     \item$\sum_{k=0}^{M-1}\sqrt{\alpha_k}\xi_k(\hat{V}^\dagger\otimes U_k)\ket{k}_a\ket{\psi} =\\
     \qquad= \ket{0}_a^{\otimes m}\ket{A\psi} + \ket{\perp}_a^{\otimes m}\ket{A^\perp\psi},$
 \end{enumerate}
 with $\ket{\perp}_a^{\otimes m}$ being a different state from $\ket{0}_a^{\otimes m}$ and $A^\perp$ being the operator complementary to $A$ so that the final global state remains a quantum normalized state.
 
Therefore, if we measure the $\ket{0}_a^{\otimes m}$ state in the ancilla qubits, we know that the rest of the qubits will have the desired state $\ket{A\psi}$. However, this is not a deterministic technique, since we only have a certain probability that we will measure in the ancilla qubits the $\ket{0}_a^{\otimes m}$ state, so it can fail.

\subsection{Hadamard+LCU Test (HoLCUs)}
Based on the above definitions, we determine how to obtain the expectation value $\bra{\psi}A\ket{\psi}$. For this, our approach is to use the Hadamard test with the LCU circuit. That is, we will have three registers: the Hadamard test ancilla register, the LCU auxiliary register and the state register. With this, we calculate the expectation value of the operator applied by the LCU protocol, since it is unitary, and we will see that its expectation value is equal to the expectation value of the non-unitary operator $A$.

\begin{figure}
\centering
\begin{tikzpicture}
\node[scale=0.8]{
\begin{quantikz}[column sep=5pt, row sep={20pt,between origins}]
\lstick{$\ket{0}_H$} \slice{1}& \gate{H}\slice{2} & \ctrl{1}\slice{3}&\ctrl{1}         & \qw            & \qw      &\qw \slice{4} & \ctrl{1}\slice{5} & \gate{H} \slice{6}       & \qw &\meter{}\\
\lstick{$\ket{0}_a$}          & \qw               & \gate[2]{V}      & \ctrl[open]{1}  & \ctrl[open]{1} & \ctrl{1} & \ctrl{1}     & \gate[2]{\hat{V}^\dagger} & \qw & \qw \\
\lstick{$\ket{0}_a$}          &\qw                &                  & \ctrl[open]{1}  & \ctrl{1}       & \ctrl[open]{1} & \ctrl{1} &                         & \qw & \qw \\
\lstick{$\ket{0}$}            &\qw                & \gate[2]{B(\ket{\psi})} & \gate[2]{U_0} & \gate[2]{U_1} & \gate[2]{U_2} & \gate[2]{U_3} & \qw             & \qw & \qw \\
\lstick{$\ket{0}$}            &\qw                &                   &                &                &          &              & \qw                       & \qw & \qw 
\end{quantikz}};
\end{tikzpicture}
\caption{Quantum circuit for the HoLCUs of a 2-qubits state $|\psi\rangle$ and an operator $A=\sum_{k=0}^{3} \alpha_k \xi_k U_k$, with two ancilla qubits for the LCU and one for the Hadamard.}
\label{cirq: H+LCU}
\end{figure}
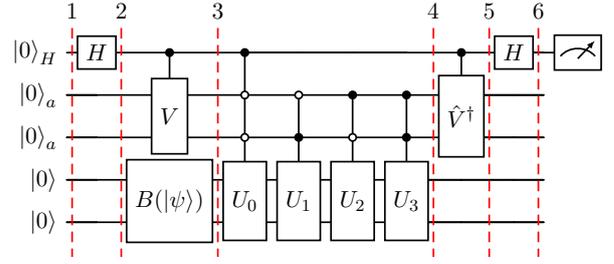

The circuit is presented in Fig. \ref{cirq: H+LCU}. The state evolves as
\begin{enumerate}
    \item$\ket{0}_H\ket{0}_a^{\otimes m} \ket{0}$
    \item$\left(\frac{\ket{0}_H+\ket{1}_H}{\sqrt{2}}\right)\ket{0}_a^{\otimes m} \ket{0}$
    \item$\frac{1}{\sqrt{2}}\left(\ket{0,0,\psi}+ \sum_{k=0}^{M-1}\sqrt{\alpha_k}\xi_k\ket{1,k,\psi}\right)$
    \item$\frac{1}{\sqrt{2}}\left(\ket{0,0,\psi} + \sum_{k=0}^{M-1}\sqrt{\alpha_k}\xi_k\ket{1,k,U_k\psi}\right)$
    \item$\frac{1}{\sqrt{2}}\left(\ket{0,0,\psi}+\ket{1,0, A\psi}+\ket{1,\perp,A^\perp\psi}\right)$
    \item$\frac{1}{2}\ket{0,0,(\mathbb{I}+A)\psi}+\frac{1}{2}\ket{0,\perp,A^\perp\psi}+\\
    +\frac{1}{2}\ket{1,0,(\mathbb{I}-A)\psi}-\frac{1}{2}\ket{1,\perp,A^\perp\psi},$
\end{enumerate}
where $\ket{x,y,z} = \ket{x}_H\ket{y}_a^{\otimes m} \ket{z}$.

So the probability $P(0)$ of measuring $\ket{0}$, or $P(1)$ of measuring $\ket{1}$, in the Hadamard qubit is
\begin{align}
    P(0)&=\frac{1}{4}\left(1+2Re[\langle A\rangle]+\langle A^\dagger A\rangle+\langle A^\perp\rangle\right),\\
    P(1)&=\frac{1}{4}\left(1-2Re[\langle A\rangle]+\langle A^\dagger A\rangle+\langle A^\perp\rangle\right),
\end{align}
so that
\begin{equation}
    Re[\bra{\psi}A\ket{\psi}] = P(0)-P(1) = 2P(0)-1,\label{eq: Real Part H+LCU}
\end{equation}
the same result as in the Hadamard test. The imaginary part is obtained in an analogous way by applying an $S^\dagger$ gate on the Hadamard qubit.

We can see that our method also only requires measuring a single qubit, the Hadamard qubit, and is deterministic, not like the LCU method, which has a probability of failure, so it is extremely convenient. As can be seen, the initialization of the LCU register can be performed in parallel to the creation of the $\ket{\psi}$ state, which allows us to optimize this step. In addition, if the coefficients of the linear combination instead of following the normalization, follow
\begin{equation}
    \sum_{i=0}^{M-1} \alpha_i = \mathcal{N}
\end{equation}
we will apply the method on an operator $A'=\frac{A}{\mathcal{N}}$ and Eq. \eqref{eq: Real Part H+LCU} becomes
\begin{align}
    &Re[\bra{\psi} A\ket{\psi}] = \mathcal{N} Re[\bra{\psi}A'\ket{\psi}] =\nonumber\\
    &= \mathcal{N}(P(0)-P(1)) = \mathcal{N}(2P(0)-1).
\end{align}

\begin{figure}
\centering
\begin{quantikz}[column sep=10pt, row sep={20pt,between origins}]
\lstick{$\ket{+}_H$} &  & \ctrl{1} &          & \ctrl{1} &          & \ctrl{1} &          & \ctrl{1} & \qw      \\
\lstick{$\ket{0}_a$} &         & \gate{H} & \swap{1} & \gate{H} & \swap{1} & \gate{H} & \swap{1} & \gate{H} & \qw      \\ 
\lstick{$\ket{0}_a$} &         &          & \targX{} & \swap{1} & \targX{} & \swap{1} & \targX{} &  \qw     & \qw      \\
\lstick{$\ket{0}_a$} &          &          &          & \targX{} & \swap{1} & \targX{} & \qw      & \qw      & \qw      \\
\lstick{$\ket{0}_a$} &          &          &          &          & \targX{} & \qw      & \qw      & \qw      & \qw      
\end{quantikz}
\caption{Quantum circuit for the uniform initialization process for the HoLCUs with four ancilla qubits.}
\label{cirq: init uniform}
\end{figure}
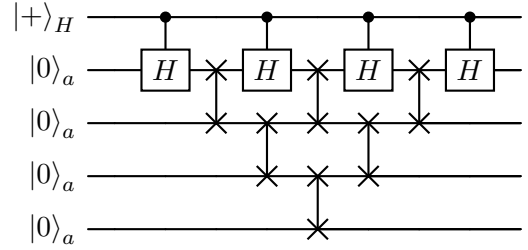

One of the difficulties with the method is the initialization of the ancilla qubits conditional on the Hadamard qubit, since it involves applying many conditional gates. However, in cases where the initialization requires few gates it can be very convenient to compute a huge number of terms at once. An example would be the case in which all the elements of the decomposition into unitary operators have the same factors $\alpha_i \xi_i$. In this case, for $M$ encoded terms in $m = \lceil \log_2(M)\rceil$ ancilla qubits, we would need to apply a conditional Hadamard gate to each qubit, which would require $m$ conditional operations. If we consider that each qubit is connected only to its nearest qubit, we need to add swaps, so that the depth would be $2m$, with $\frac{m(m+1)}{2}$ operations. We can see an example in Fig. \ref{cirq: init uniform} with four ancilla qubits.

This can be useful in decompositions in which we have a small number of different $\alpha_i \xi_i$ factors repeated multiple times, so that we can compute at the same time with a single circuit each group of unitary operators.

\begin{table*}
    \centering
    \begin{tabular}{|l|l|l|l|l|}
         \hline
         & Number of qubits & Circuit Depth & Circuits & Shots \\
         \hline
        Raw & $n$& $d_0$ & 1  & $O(2^n/\varepsilon)$\\
        Hadamard Test& $n+1$ & $d_0+d_u$ & $M$ & $O(1/\varepsilon^2)$\\
         \hline
        HoLCUs & $n+\lceil \log_2(M)\rceil+1 $ & $O(d_0+d_uM+d_m)$ & 1 & $O(1/\varepsilon^2)$\\
        HoLCUs (div) & $n+\lceil \log_2(M/M_t)\rceil+1$ & $O(d_0+d_uM/M_t+m'^2)$ & $M_t$ & $O(1/\varepsilon^2)$\\
         \hline
    \end{tabular}
    \caption{Comparison between methods for estimating the expectation value. $n$ is the number of state $\ket{\psi}$ qubits, $M$ the number of LCU terms, $M_t$ the number of groups of terms with different coefficients in the LCU, $m = \lceil \log_2(M)\rceil$, $m' = \lceil \log_2(M/M_t)\rceil$, $d_0$ the depth for $\ket{\psi}$ state preparation, $d_m$ the depth to initialize a LCU state of $m$ qubits, $d_u$ the depth for the controlled unitary application and $\varepsilon$ the estimation error.}
    \label{tab: Expectation}
\end{table*}

We compare in Tab. \ref{tab: Expectation} the characteristics of the different methods. These are:
\begin{itemize}
    \item Raw: Obtain the expectation value by sampling the state directly and assigning the expectation value to each combination obtained.
    \item Hadamard: Apply the Hadamard test to each unitary operator of the linear combination and sum the weighted expectation values.
    \item HoLCUs: Apply the HoLCUs method to the complete operator $A$.
    \item HoLCUs (div): Group the unitary operators that have the same factor in the linear combination of the global operator and calculate the expectation value with the uniform HoLCUs explained above.
\end{itemize}

Let us compare for a $\ket{\psi}$ state of $n$ qubits, with an operator A with $M$ terms and $M_t$ different factor groups, each of size $M_i$ terms. In addition, $m = \lceil \log_2(M)\rceil$ and $m_i = \lceil \log_2(M_i)\rceil$. To consider a simple case, we choose that all groups are of the same size, which we call $M_0=M/M_t$.

We need $d_0$ depth to prepare the $\ket{\psi}$ state, which in a QAOA will involve $O(M^2)$ operations and $d_m$ depth to initialize a LCU state of $m$ qubits, which will be $O(2^m)=O(M)$. We assume that our factors satisfy the normalization condition to simplify the analysis. We consider that the depth of application of each conditional unitary is $d_u$.

Another important point is that, as depicted in Fig.~\ref{cirq: H+LCU}, in general the Hadamard qubit only applies a controlled operation on the ancilla qubits of the LCU and not to the state qubits, except in the $U_0$ term. This is because the $U_0$ term is the only term that would be activated for state $\ket{0}_H$ in the Hadamard qubit. In case we have a number of terms less than $2^n$, we can shift the initialization so that the free space we have left is in the $\ket{0}^{\otimes n}_a$ state of the LCU qubits. This way we avoid a multicontrol gate that relates at the same time the Hadamard qubit, the ancilla qubits and the state qubits. This is because the value $\ket{0}^{\otimes n}_a$ in the control qubits will only be given in the case of not having applied the $V$ operator, since it will not initialize the $\ket{0}^{\otimes n}_a$, and all states other than $\ket{0}^{\otimes n}_a$ are only activated when $V$ is applied.

The scaling in precision when obtaining the expectation value is the same as in the standard Hadamard Test, since we have the same final equation and we only need to accurately determine the probability of obtaining 0 or 1 by measuring one qubit in one circuit.

\section{Experiments design}
In order to study its potential, we have applied the HoLCUs Test method to the QAOA algorithm. The objective is to reduce the number of runs and shots of the circuit to obtain the expectation values of the hamiltonian of the system, and then, the training time. As our hamiltonian is a hermitian operator, we only need to apply the circuit to calculate the real part of the expectation value.

We have applied the method to calculate all the terms at the same time, without taking advantage of the repetition mentioned above. We have done this in order to check the performance of the method applied in the more general case without taking advantage of the concrete problem. We also compare the HoLCUs QAOA against an simple QAOA with Hadamard test to avoid hidden optimizations of another implementations, having a better comparative. We have performed two tests for solving QUBO problems, with random cost matrices with values in range $(-2,2)$.

In the first test we will compare the run time of the usual QAOA with Hadamard  Test and the HoLCUs QAOA. This test consists in running the QAOA training algorithm with 1, 2 and 3 layers, for each number of variables $n$ from 3 to 9. In each experiment we use $10^4$ shots for the expectation value calculation and 3 different initializations of the QAOA parameters. We test with different 5 instances per $n$. We implemented the circuit in Qiskit and simulated it with the AerSimulator with GPU, max\_parallel\_shots=0 and max\_parallel\_threads=0. The execution is performed in Google Colab with GPU.

In the second test we will study the scaling of the run time of the HoLCUs QAOA algorithm in more detail. In this case, we go from 3 to 11 variables, with 10 instances per $n$ and 3 layers. The other hyperparameters are the same. This test is performed in Google Colab with CPU.

In both experiments the run time considered includes all the processes after transforming the QUBO into the Ising. So, it includes the circuit creation, transpilation and optimization, which has higher run time in the HoLCUs case, and the optimization of the angles, with the expectation value computation, which has higher run time in the Hadamard case.

\section{Results}
\begin{figure}
    \centering
    \includegraphics[width=\linewidth]{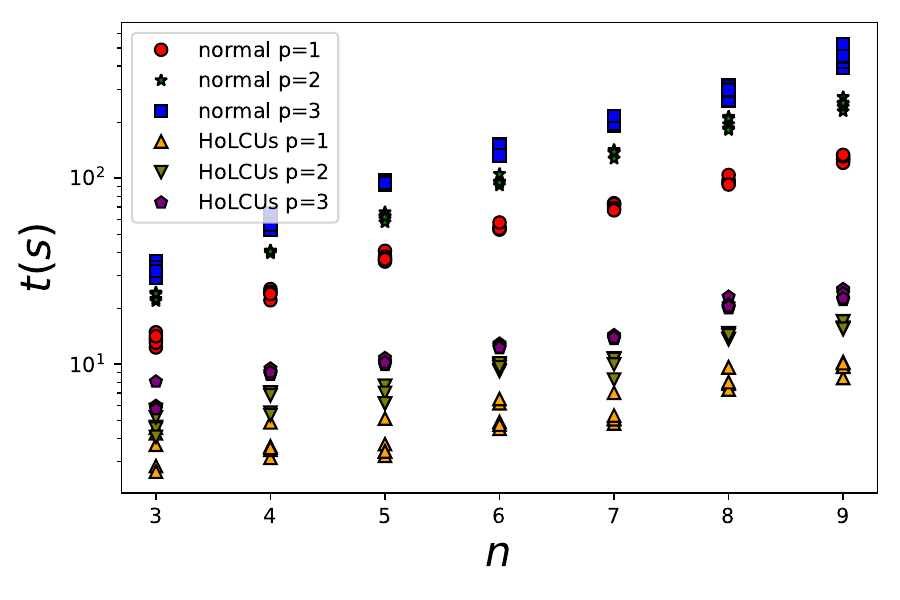}
    \caption{Run time vs number of variables $n$ for $p=1,2,3$. Comparative of Hadamard Test QAOA and HoLCUs QAOA.}
    \label{fig: Time vs n}
\end{figure}

In Fig.~\ref{fig: Time vs n} we can observe the run time of the QAOA algorithm with the Hadamard Test and with the HoLCUs versus the number of variables $n$. We run a series of instances for each value of the number of layers $p$ from 1 to 3. The HoLCUs QAOA has a lower run time for the same instances, as well as a lower run time growth. We can also see that both algorithms grow in run time with the number of layers and with the number of variables. However, the HoLCUs QAOA runtime is always below the normal QAOA run time.

\begin{figure}
    \centering
    \includegraphics[width=\linewidth]{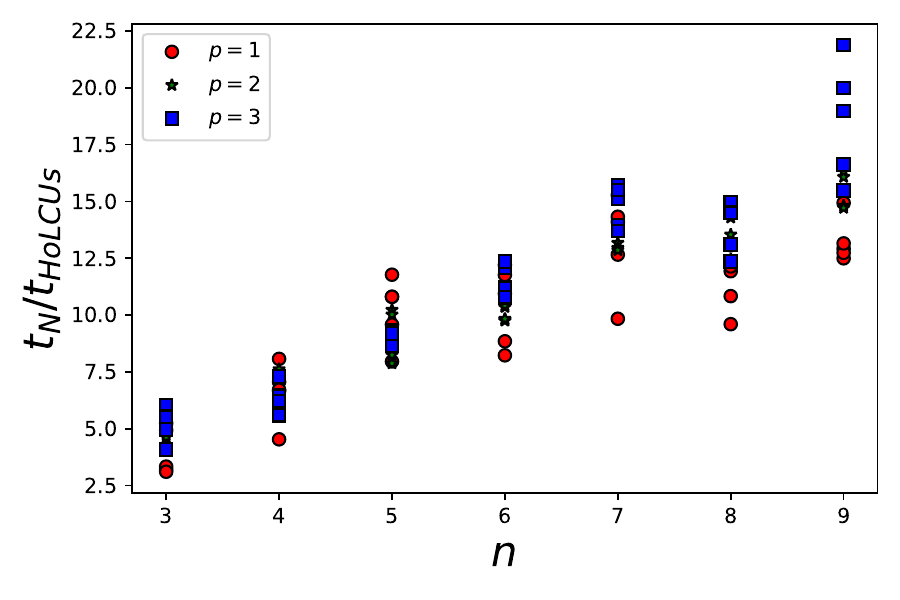}
    \caption{Speedup $t_N/t_{HoLCUs}$ vs number of variables $n$ for $p=1,2,3$.}
    \label{fig: Time vs n Accel}
\end{figure}

We can observe in Fig.~\ref{fig: Time vs n Accel} that the HoLCUs QAOA is between $2.5$ and $22.5$ times faster than the QAOA with Hadamard Test in the tests performed, increasing with problem size in a trend that appears linear. The increase in speedup was expected due to the increasing number of terms of the Ising problem with the increase of the number of variables. However, even if the number of terms increase as $O(n^2)$, the speedup does not increase that fast due to the growth of the extra ancilla qubits. We can also observe that the dispersion of the speedups also increases with the number of variables.

\begin{figure}
    \centering
    \includegraphics[width=\linewidth]{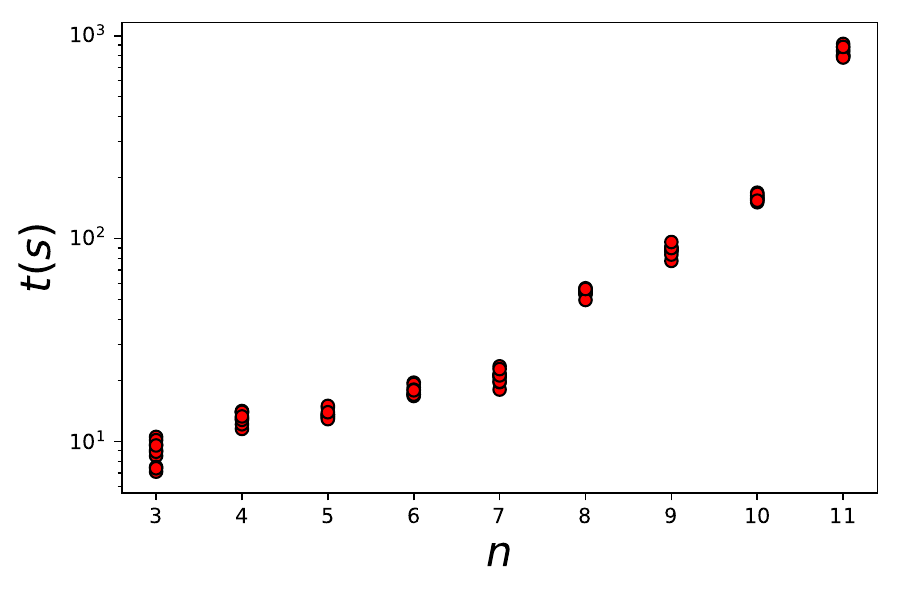}
    \caption{Run time vs number of variables $n$ for $p=3$ for the HoLCUs QAOA.}
    \label{fig: Time vs n HoLCUs}
\end{figure}

In Fig.~\ref{fig: Time vs n HoLCUs} we can observe that the execution time of the HoLCUs QAOA grows more than exponentially with the problem size addressed, reaching 1000 seconds in the case of $n=11$. This possibly is due to the extra logarithmic qubits added to the circuit, increasing the execution time of each circuit more than usually. This scaling could be improved by optimising the creation and transpilation of the circuit.

\section{Conclusions}
We have presented an algorithm that allows to determine the expectation value of an operator in an efficient way by combining the Hadamard Test and the LCU protocol. In addition, we have studied several possible optimizations to it. We have also applied it to the QAOA solving for QUBO problems, and we have seen its speedup in Qiskit simulations.

Next lines of research may include how to optimize the application of the LCU to the state, since it is the most expensive part at the depth level, which types of QUBO problems are more susceptible to use this method or how to adapt it to error susceptible devices. It also open a way to make unitary matrix product operators in the tensor networks approaches to compute expected values, allowing to exploit new properties.

\section*{Acknowledgement}
This work has been developed in the `When Physics Becomes Science' project of \href{https://www.youtube.com/@whenphysics}{When Physics}, an initiative to recover the original vision of science.

\bibliographystyle{IEEEtran}
\bibliography{biblo}
\end{document}